\documentclass[apj]{emulateapj}

\usepackage{mathtools}

\usepackage{times}
\usepackage{multirow}
\usepackage{amssymb}
\usepackage{amsmath}
\usepackage{pifont}
\usepackage{graphics}

\usepackage[svgnames]{xcolor}
\usepackage{epsfig}
\usepackage{threeparttable}

\usepackage[normalem]{ulem}
\usepackage{color}

\newcommand{\MS}{M\textsubscript{$\odot$}}    
\newcommand{\RS}{R\textsubscript{$\odot$}}    
\newcommand{\ME}{M\textsubscript{$\oplus$}}   
\newcommand{\RE}{R\textsubscript{$\oplus$}}   

\newcommand{\bjdtdb}{\ensuremath{\rm {BJD_{TDB}}}}

\begin{document}

\title{A Multi-year Search for Transits of Proxima Centauri. \\II: No evidence for transit events with periods between 1--30 days}

\author{Dax L. Feliz\altaffilmark{1,2,3}, David L. Blank\altaffilmark{4}, Karen A. Collins\altaffilmark{5}, Graeme L. White\altaffilmark{4}, Keivan G.\ Stassun\altaffilmark{1,2}
, Ivan A. Curtis\altaffilmark{6}, Rhodes Hart\altaffilmark{4}, John F. Kielkopf\altaffilmark{7}, Peter Nelson\altaffilmark{8}, Howard Relles\altaffilmark{5},
Christopher Stockdale\altaffilmark{9}, Bandupriya Jayawardene\altaffilmark{4}, 
Paul Shankland\altaffilmark{10}, 
Daniel E. Reichart\altaffilmark{11}, Joshua B. Haislip\altaffilmark{11}, and Vladimir V. Kouprianov\altaffilmark{11}}

\altaffiltext{1}{Department of Physics and Astronomy, Vanderbilt University, Nashville, TN 37235, USA}
\altaffiltext{2}{Department of Physics, Fisk University, 1000 17th Avenue North, Nashville, TN 37208, USA}
\altaffiltext{3}{Corresponding Author, dax.feliz@vanderbilt.edu}
\altaffiltext{4}{Centre for Astrophysics, University of Southern Queensland, Toowoomba, Queensland 4350, Australia}
\altaffiltext{5}{Harvard-Smithsonian Center for Astrophysics, Cambridge, MA 02138, USA}
\altaffiltext{6}{ICO, Adelaide, South Australia}
\altaffiltext{7}{Department of Physics and Astronomy, University of Louisville, Louisville, KY 40292, USA}
\altaffiltext{8}{Ellinbank Observatory, Victoria, Australia}
\altaffiltext{9}{Hazelwood Observatory, Churchill, Victoria, Australia}
\altaffiltext{10}{U.S. Naval Observatory, Flagstaff Station, 10391 W Naval Observatory Rd, Flagstaff, AZ 86001, USA}
\altaffiltext{11}{Department of Physics and Astronomy, University of North Carolina at Chapel Hill, Campus Box 3255, Chapel Hill, NC 27599, USA}



\begin{abstract}
Using a global network of small telescopes, we have obtained light curves of Proxima Centauri at 329 observation epochs from 2006 -- 2017. The planet Proxima~b discovered by \citet{Anglada:2016} with an orbital period of 11.186~d has an \textit{a priori} transit probability of $\sim 1.5\%$; if it transits, the predicted transit depth is about 5 millimagnitudes. In \citet{Blank:2018}, we analyzed 96 of our light curves that overlapped with predicted transit ephemerides from previously published tentative transit detections, and found no evidence in our data that would corroborate claims of transits with a period of 11.186~d. Here we broaden our analysis, using 262 high-quality light curves from our data set to search for any periodic transit-like events over a range of periods from 1 -- 30~d. We also inject a series of simulated planet transits and find that our data are sufficiently sensitive to have detected transits of 5 millimagnitude depth, with recoverability ranging from $\sim$100\% for an orbital period of 1~d to $\sim$20\% for an orbital period of 20~d for the parameter spaces tested. Specifically at the 11.186~d period and 5~millimagnitude transit depth, we rule out transits in our data with high confidence. We are able to rule out virtually all transits of other planets at periods shorter than 5 d and depths greater than 3 millimagnitudes; however, we cannot confidently rule out transits at the period of Proxima b due to incomplete orbital phase coverage and a lack of sensitivity to transits shallower than 4 millimagnitudes. 
\end{abstract}

\section{Introduction}\label{sec:intro}
The discovery of Proxima Centauri~b (Proxima~b, hereafter) via the radial velocity (RV) technique by \citet{Anglada:2016} was a landmark event in exoplanet studies. We now know that orbiting in the habitable zone \citep{Kopparapu:2013} of the star nearest to our Sun is a planet that is likely to be rocky 
\citep{Brugger:2016,Kane:2017, Bixel:2017} and possibly habitable \citep{Ribas:2016,Barnes:2016,Meadows:2016,Turbet:2016,Boutle:2017}. We report here further results from our transit search of Proxima Centauri from 2006 to 2017 \citep[Paper I hereafter]{Blank:2018} which was motivated by the possibility that such planets may exist, and that they could be found using sub-meter size telescopes with commercial grade CCD cameras.

\setcounter{footnote}{0}
In the 11 years of this photometric campaign, we collected light curves at 329 epochs. Of these 329 light curves, 262 passed various quality tests (detailed in Paper I), 96 of which overlapped with the previously published ephemerides\footnote{Throughout this paper we use the word ephemeris to refer to predicted, known, or estimated reference transit center time $\mathrm{T_c}$ plus an orbital period P for a known or possible transiting exoplanet. These values can be derived precisely from transit observations, or with less precision from RV observations.} of \citealt{Anglada:2016}, \citealt{Kipping:2017}, \citealt{Liu:2017} and \citealt{Li:2017}. A search for transits corresponding to Proxima~b at these ephemerides is reported in Paper~I. No convincing transit event attributable to Proxima~b was detected in this subset of light curves. 

In this work, we proceed to search all 262 quality light curves systematically for a planet of any orbital period in the period range 1.01 to 30.5~days. In Section \ref{sec:datareduction} we summarize the data collection, drawing reference to Paper~I. We also describe our strategy for determining a period range to conduct our planet search. In Section \ref{sec:methodology} we describe our methods of analysis for searching for periodic transit events and tests for statistical significance and sensitivity to detect transit events. Section \ref{sec:results} contains our results and the tests of sensitivity needed to place limits on any possible detection. We discuss our findings in Section \ref{sec:discussion}.

\section{Observations and Data Completeness}\label{sec:datareduction}
\subsection{Summary of Observations and Data Reduction done in \citet{Blank:2018}}
The observations that make up our data set of 329 light curves came from the world-wide robotic telescope network Skynet \citep{Reichart:2005}, the Real Astronomy Experience (RAE) robotic telescope \citep{Fadavi:2006} located in in Bickley, Western Australia, and from several participating observatories from the Kilodegree Extremely Little Telescope (KELT; \citealt{Pepper:2007,Pepper:2012}) Follow-Up Network (KELT-FUN; \citealt{Collins:2018}). More details about the participating instruments are in Table 2 in Paper I.

Our data reduction techniques are described in Section 3.2 of Paper I. To briefly summarize, in order to minimize the effects of long term stellar variability and differential chromatic airmass we performed a linear detrend for each parameter so that the final mean flux value was 1.0. Additionally, for correlated changes in the photometric baseline due to telescope meridian flips, we fitted and realigned the baseline at that point. In some light curves we performed an additional set of linear detrends using the x- and/or y-centroid locations of the target star, sky background, full-width half-maximum of the stellar point spread function, and/or the total number of comparison star net integrated counts. For more information on detrending with AstroImage J, see Section 4.4 of \citet{Collins:2017}. To remove obvious flares that are predicted to occur $\sim63$ times per day \citep{Davenport:2016} and photometric outliers, we performed an iterative 3-$\sigma$ clipping. We present our 3-$\sigma$ clipped undetrended and detrended light curves in Tables \ref{tbl:data1} and \ref{tbl:data2} in the Appendix Section.

\subsection{Geometric Transit Probability of Proxima b-Like Planets For A Given Period\label{GTP}} 
To estimate the geometric transit probability of Proxima, we assumed a planet mass of 1.27 \ME ~and a planet radius of $\sqrt{\delta} R_*$, where $\delta \sim $ 5 millimagnitudes (or mmag, hereafter) as reported by \citealt{Anglada:2016} and then calculated the probability of a planet transiting as the fraction of the area of the celestial sphere that is swept out by the shadow of the planet during one orbital period \citep{Borucki:1984,Winn:2010}: 

\begin{equation}\label{eq:GeoTransProb}
     \text{Transit Probability} =
     \big(\frac{R_* + R_p}{a}\big)\big(\frac{1+esinw}{1-e^2} \big)
\end{equation}
Where R$_*$ is the stellar radius, R$_p$ is the planet radius, a is the semi-major axis, e is the eccentricity and w is the argument of periastron. Applying Kepler's Third Law, assuming e=0, w=$\pi/2$ and that $R_p \ll R_*$ gives
\begin{equation}
\begin{aligned}
    \frac{R_*}{a}= R_* \big(\frac{G(M_p+M_*)P^2}{4\pi^2}\big)^{-1/3} 
\end{aligned}
\end{equation}

\noindent
Adopting a stellar mass M$_*$ = 0.1221 \MS~and stellar radius R$_*$ =  0.1542 \RS~\citep{Kervella:2017}, the geometric probability of transit detection for orbital periods from 0.01 to 365 days is shown in the top panel of Figure \ref{fig:phase_coverage}. It should be noted that this geometric estimation is based on the planet density model assumed by \citealt{Anglada:2016} ($M_p= 1.27~ \ME, \rho/ \rho_{\Earth}$ = 1) and that all transits are assumed to be across the face of the host star. Scaled curves of this type will have varied results from different assumptions of the density of the planet and from grazing transits.

\subsection{Phase Coverage of Photometric Observations\label{sec:Phase_Coverage}}
To estimate the phase coverage of our data, we phase folded our data for each day in the period range of 0.01 -- 365 days and then binned our data into 5 minute bins and calculated the inverse variance weighted means for each bin. 
\begin{equation}
    \Hat{y} = \frac{\sum_i y_i / \sigma_i^2}{\sum_i 1/\sigma_i^2}
\end{equation}

\begin{figure}[htp] 
\centering 
\includegraphics[width=\columnwidth]{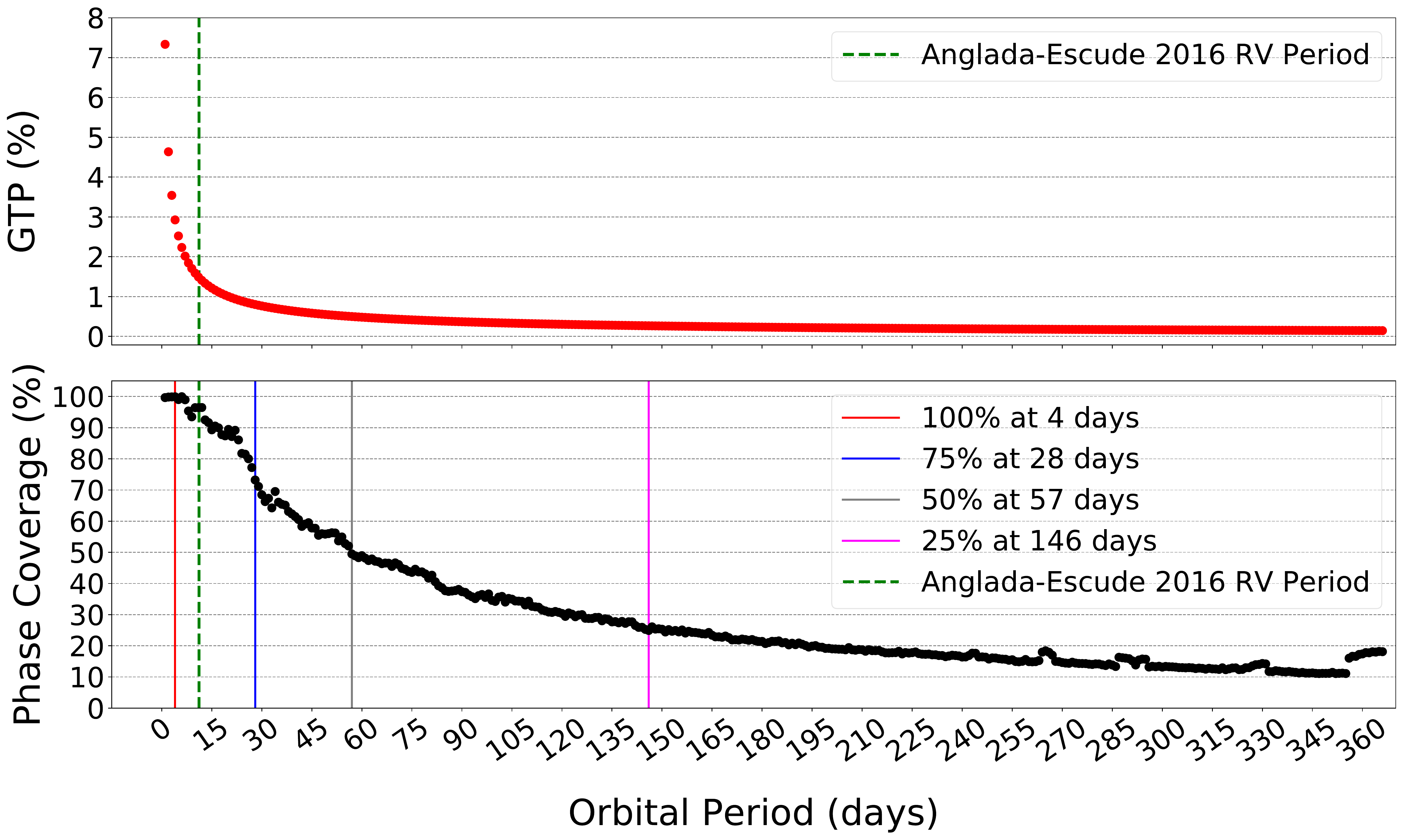}
\caption{ Top: The Geometric Transit Probability (GTP) of Proxima Centauri. At the orbital period of 11.186 days determined by the radial velocity discovery of \citet{Anglada:2016} the transit probability is $\sim 1.5 \%$ based on their assumed planet density model. Bottom: Phase Coverage of the 329 individual light curves phase folded and binned into 5 minute bins, as described in Section \ref{sec:Phase_Coverage}. The colored solid lines mark the periods where phase coverage is 100\%, 75\%, 50\% and 25\%. The phase coverage of our data drops below $\sim 75\%$ around 30 days.}
\label{fig:phase_coverage}
\end{figure}

\noindent
We then define phase coverage as the number of finite values in our phase folded data bins divided by the total number of values in our phase folded data bins. The result of this procedure is shown in the bottom panel of Figure \ref{fig:phase_coverage} where it is clear that the phase coverage falls below $\sim75\%$ for periods longer than 30 days. Both panels of Figure 1 show the increasing difficulty of detection of a transit event for longer periods. For period longer than 30 days, both the low sensitivity of detection and the poor phase coverage imply a low probability of detecting a transit, and we have decided to limit our search to the period region 1.01 to 30.5 days. A lower orbital period limit of 1.01 days was chosen to avoid potential aliases due to diurnal and sidereal day sampling.

\section{Methodology}\label{sec:methodology}

\subsection{Box-fitting Least-Squares Algorithm }\label{sec:BLS}
The box-fitting least-squares (BLS) algorithm \citep{Kovacs:2002}  searches for periodic decreases in star brightness of a photometric time series. The BLS algorithm models transit events as simple step functions and identifies transiting planet candidates by phase folding light curves to trial frequencies and searching a grid of transit epochs and durations at each frequency, and then picking the parameters that maximize the transit
depth significance with a least-squares optimization. We use the VARTOOLS software \citep{Hartman:2016} to produce the BLS power spectrum shown in Figure \ref{fig:BLS_1_30}. To determine the fractional transit length (transit duration divided by orbital period) $q$, we estimated a minimum value $q_{min}$ = 0.017 to account for transit of at least 25 minutes duration for a minimum orbital period of 1.01 days, and $q_{max}$ = 0.1 for events of $\sim 3$ days for a maximum orbital period of 30.5 days. With $q_{min}$, a range of desired frequencies ($f_{max}=1/P_{min}$, $f_{min}=1/P_{max}$), 
and the total cumulative baseline of our full data set (T) , we can roughly estimate the number of frequencies required for our BLS search:
 
\begin{equation}\label{eq:Nfreq}
      N_{freq} = {(f_{max} - f_{min})}/df
\end{equation}
where the stepsize\footnote{https://www.astro.princeton.edu/~jhartman/vartools.html\#BLS}\label{ftn:vartools}, df is defined as:
\begin{equation}\label{eq:df}
      df = q_{min}/4T
\end{equation}

To estimate the number of phase bins\footnotemark[2] ($N_{bin}$) to break up our cumulative light curve into, we used $2/q_{min} \sim$ 120 bins. All parameters used in the BLS search are listed in Table \ref{tbl:BLSparam}. \\

\begin{table*}[htb!]
\begin{center}
\caption{Parameters used in BLS periodic searches
\label{tbl:BLSparam}}{\setlength{\tabcolsep}{0.30em}
\begin{tabular}{lcccccc}
\tableline
\multicolumn{1}{l}{$q_{min}$} & \multicolumn{1}{c}{$q_{max}$} & \multicolumn{1}{c}{$P_{min}~(days)$} & \multicolumn{1}{c}{$P_{max}~(days)$} & \multicolumn{1}{c}{$N_{freq}$} & \multicolumn{1}{c}{$N_{bin}$} &
\multicolumn{1}{c}{$df~(1/seconds)$}\\ 
\vspace{-0.1in}\\
\tableline
\vspace{-0.1in}\\
0.017	& 0.1 & 1.01 & 30.5 & 920,492  &  120 & $1.05\times10^{-6}$\\
\end{tabular}
}
\end{center}
\end{table*}

\subsection{False Alarm Probability\label{sec:FAP}}
To determine the statistical significance of peaks in the BLS power spectrum, we define a false alarm probability (FAP) to be the probability of a peak having equal strength by random chance or due to the cadence of our sampling. This is done by randomly rearranging the detrended fluxes and error information while keeping the time stamps fixed, reapplying the BLS search and recording the BLS outputs for the top peak of each iteration. This random permutation is then repeated 1,000 times. The 0.1\% FAP is the highest peak out of 1,000 permutations, the 1\% FAP is the 10th highest peak out of 1,000 permutations and the 10\% FAP is the 100th highest peak out of 1,000 permutations. In Section \ref{sec:FAP_tests}, we apply different variations of this definition of FAP.

\subsection{Transit Injections\label{sec:TransitInjections}}
To test the sensitivity of the BLS algorithm in recovering transit-like events, we injected fake transits into our detrended data and ran the BLS algorithm on the injected data sets. We simulated these fake transits using a given transit depth $\delta$, and orbital periods $P$, along with the stellar mass, stellar radius, the orbital eccentricity \textit{e}, orbital inclination \textit{i} and the argument of periastron \textit{w} (all transit model parameters are listed in Table \ref{tbl:TransInjParam}). With each permutation of our transit model parameters, we simulated a total of 550 Mandel-Agol transit models \citep{Mandel:2002} using the Python package PyTransit, \citep{Parviainen:2015}, which in addition to the parameters listed in Table \ref{tbl:TransInjParam}, also uses quadratic limb darkening coefficients, \textit{u}. To obtain values for \textit{u}, we used the EXOFAST\footnote{http://astroutils.astronomy.ohio-state.edu/exofast/limbdark.shtml\label{ftn:exofast}}\citep{Eastman:2013} website to interpolate quadratic limb darkening coefficients from the limb darkening tables in \citet{Claret:2011}. As seen in Table 2 of Paper I, the majority of our light curves were observed with an R filter and we were able to obtain u $\sim$[0.425, 0.298] for Proxima in the R band using the EXOFAST website, providing $\mathrm{T_{Eff}}$ = 3042 K, log g = 5.20, [Fe/H] = 0.21 \citep{Segransan:2002} and as inputs.  After model creation, we then use BLS to test our ability to successfully detect our injected transit events, as described in Section \ref{sec:TrasitInjectionRecovery}. 

\begin{table*}[htb!]
\begin{center}
\caption{Parameters used in Transit injection Analysis\label{tbl:TransInjParam}}{\setlength{\tabcolsep}{0.30em}
\begin{tabular}{lcc}
\tableline

\multicolumn{1}{|c|}{Parameter}& \multicolumn{1}{c}{Value / Model}&\multicolumn{1}{|c|}{Citation}    \\ \hline            

\multicolumn{1}{|c|}{Stellar Radius, $\mathrm{R_{star}}$} & \multicolumn{1}{c|}{0.1542 \RS}&\multicolumn{1}{|c|}{\citet{Kervella:2017}}    \\ \hline

\multicolumn{1}{|c|}{Stellar Mass, $\mathrm{M_{star}}$} & \multicolumn{1}{c|}{0.1221 \MS}&\multicolumn{1}{|c|}{\citet{Kervella:2017}}    \\ \hline

\multicolumn{1}{|c|}{Effective Temperature,  $\mathrm{T_{Eff}}$} & \multicolumn{1}{c|}{3042 K}&\multicolumn{1}{|c|}{\citet{Segransan:2002}}    \\ \hline

\multicolumn{1}{|c|}{log g} & \multicolumn{1}{c|}{5.20}&\multicolumn{1}{|c|}{\citet{Segransan:2002}}    \\ \hline

\multicolumn{1}{|c|}{[Fe/H]} & \multicolumn{1}{c|}{0.21}&\multicolumn{1}{|c|}{\citet{Schlaufman:2010}}    \\ \hline

\multicolumn{1}{|c|}{Transit Depth, $\delta$ (mmag)}                   & \multicolumn{1}{c|}{1.0, 2.0, 3.0, 4.0, 5.0, 6.0, 7.5, 10.0,  15.0, 20.0}&\multicolumn{1}{|c|}{This work}    \\ \hline            

\multicolumn{1}{|c|}{Orbital Period, P (days)}                   & \multicolumn{1}{c|}{1.1, 2.1, 3.1, 5.1, 7.6, 10.1, 11.186, 15.1, 20.1, 25.1, 30.1}&\multicolumn{1}{|c|}{This work}    \\ \hline            

\multicolumn{1}{|c|}{Orbital Phase}                   & \multicolumn{1}{c|}{-0.4, -0.2, 0.0, 0.2, 0.4}&\multicolumn{1}{|c|}{This work}    \\ \hline          

\multicolumn{1}{|c|}{Planetary Radius, $\mathrm{ R_{planet}}$} & \multicolumn{1}{c|}{$ R_{planet} \approx \sqrt{\delta}R_{star}$}&\multicolumn{1}{|c|}{This work}     \\ \hline

\multicolumn{1}{|c|}{Eccentricity, e} & \multicolumn{1}{c|}{0.0}&\multicolumn{1}{|c|}{This work}     \\ \hline

\multicolumn{1}{|c|}{Inclination, i} & \multicolumn{1}{c|}{$\pi/2$}&\multicolumn{1}{|c|}{This work}     \\ \hline

\multicolumn{1}{|c|}{Argument of Periastron, w} & \multicolumn{1}{c|}{$\pi/2$}&\multicolumn{1}{|c|}{This work}     \\ \hline

\multicolumn{1}{|c|}{Quadratic Limb Darkening Coefficients} & \multicolumn{1}{c|}{(0.425, 0.298) for R band}&\multicolumn{1}{|c|}{Interpolated from \citet{Claret:2011} tables.}      \\ \hline
\end{tabular}}
\vspace{-0.1in}
\footnotetext[0]{\footnotesize{NOTE: The quadratic limb darkening coefficients are estimated using the EXOFAST website\textsuperscript{3} to interpolate the quadratic limb darkening \\ \hspace*{0.075in} tables from \citet{Claret:2011} by providing log g, [Fe/H] and $\mathrm{T_{Eff}}$ as inputs.}}
\end{center}
\end{table*}

\section{Results}\label{sec:results}

\subsection{BLS Power Spectrum} \label{sec:PowerSpectrum}
Using the parameters from Table \ref{tbl:BLSparam}, we applied the VARTOOLS \citep{Hartman:2016} BLS algorithm on our combined and detrended 262 observations from the Skynet, KELT-FUN and RAE telescopes. In our application of BLS, we used the ``nobinnedrms" option in VARTOOLS which calculates the Signal Residue, \textit{SR(f)}, as defined in \citealt{Kovacs:2002}, with the average value  of \textit{SR(f)} subtracted, and divided by the standard deviation of \textit{SR(f)}. This leads to points in the power spectrum that have \textit{SR(f)} below the average value and will have a negative Spectroscopic Signal to Noise (\textit{S/N(f)}, described in Section \ref{sec:FAP_2}). We ran our BLS search from 1.01 to 30.5 days and found that all peaks in the power spectrum within orbital periods of 1.01 -- 30.5 days fall below the majority of the calculated FAP and FAP(P) thresholds, as described in Section \ref{sec:FAP_tests}. The top peak of the BLS power spectrum corresponds to an orbital period of $\sim 1.808$ days, which lies above the 10\% FAP(P) threshold. As an example of the transit-like events detected by BLS in our data, Figure \ref{fig:BLS_1_30_top_peak} displays the phase folded light curve of our data, using the orbital period and transit center time reported by the BLS algorithm. Figure \ref{fig:PFLCs_topBLSPeak} in the Appendix section, shows the 32 individual light curves that contribute to this signal. Although there is some evidence for a transit-like event in Figure \ref{fig:BLS_1_30_top_peak}, the existence of such an event is not supported by analysis of the individual light curves.

\begin{figure*}[htp] 
\includegraphics[width=\linewidth, trim=0cm 0.0cm 0cm 0.0cm, clip=true]{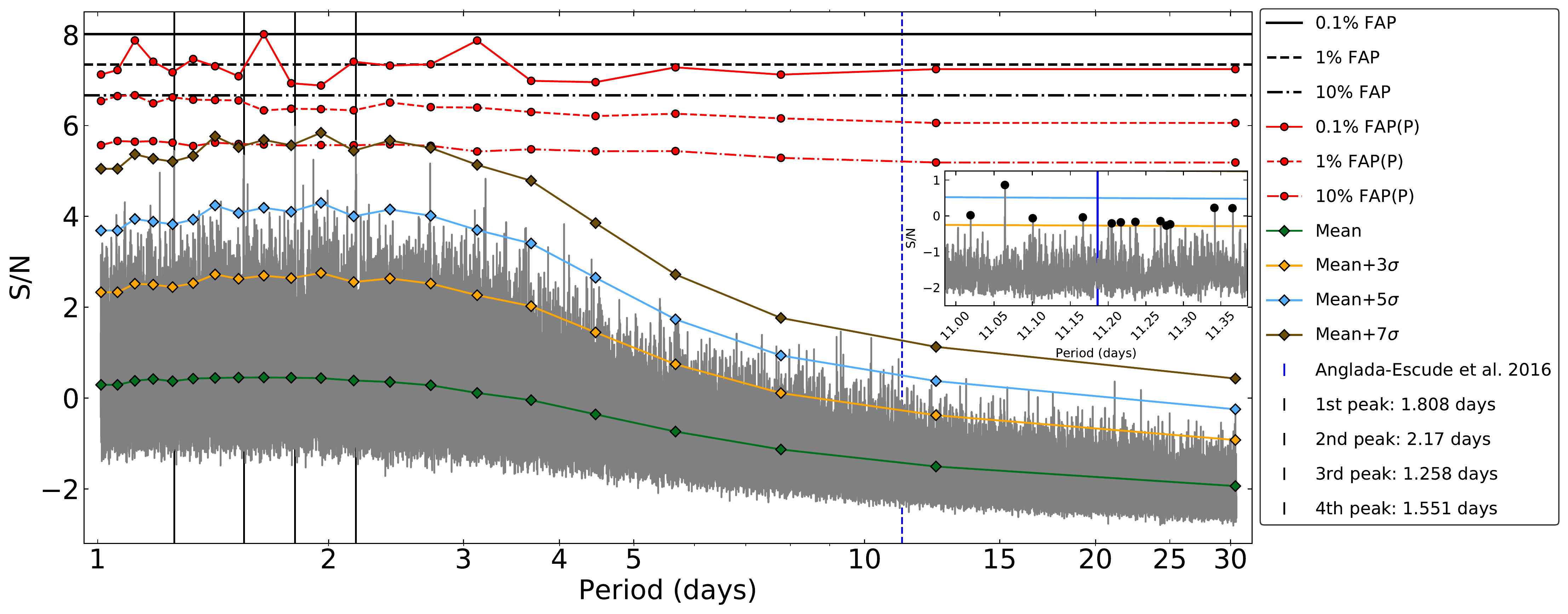}

\caption{We present a power spectrum from the VARTOOLS BLS transit search algorithm. The black vertical solid lines represent the orbital periods corresponding to the top 4 peaks of the power spectrum. The horizontal black lines correspond to the 0.1\%, 1\%, and 10\% FAP as described in Section \ref{sec:FAP}. The red lines are the FAP(P) thresholds calculated in 20 period ranges with equal $N_{freq}$ as described in Section \ref{sec:FAP_P}. The green, orange, cyan and brown lines represent the robust estimations for the mean and mean plus the 3$\sigma$, 5$\sigma$ and 7$\sigma$ of the S/N in each period range as described in Section \ref{sec:Huber}. The inset figure is a close up of peaks in the power spectrum that are near the 11.186 day RV period and are also above the mean + $3\sigma$ and mean + $5\sigma$ lines, marked with black dots. }
\label{fig:BLS_1_30}
\end{figure*}

\begin{figure}[htp] 
\centering \includegraphics[width=\columnwidth, trim=0cm 0.0cm 0cm 0.0cm, clip=true]{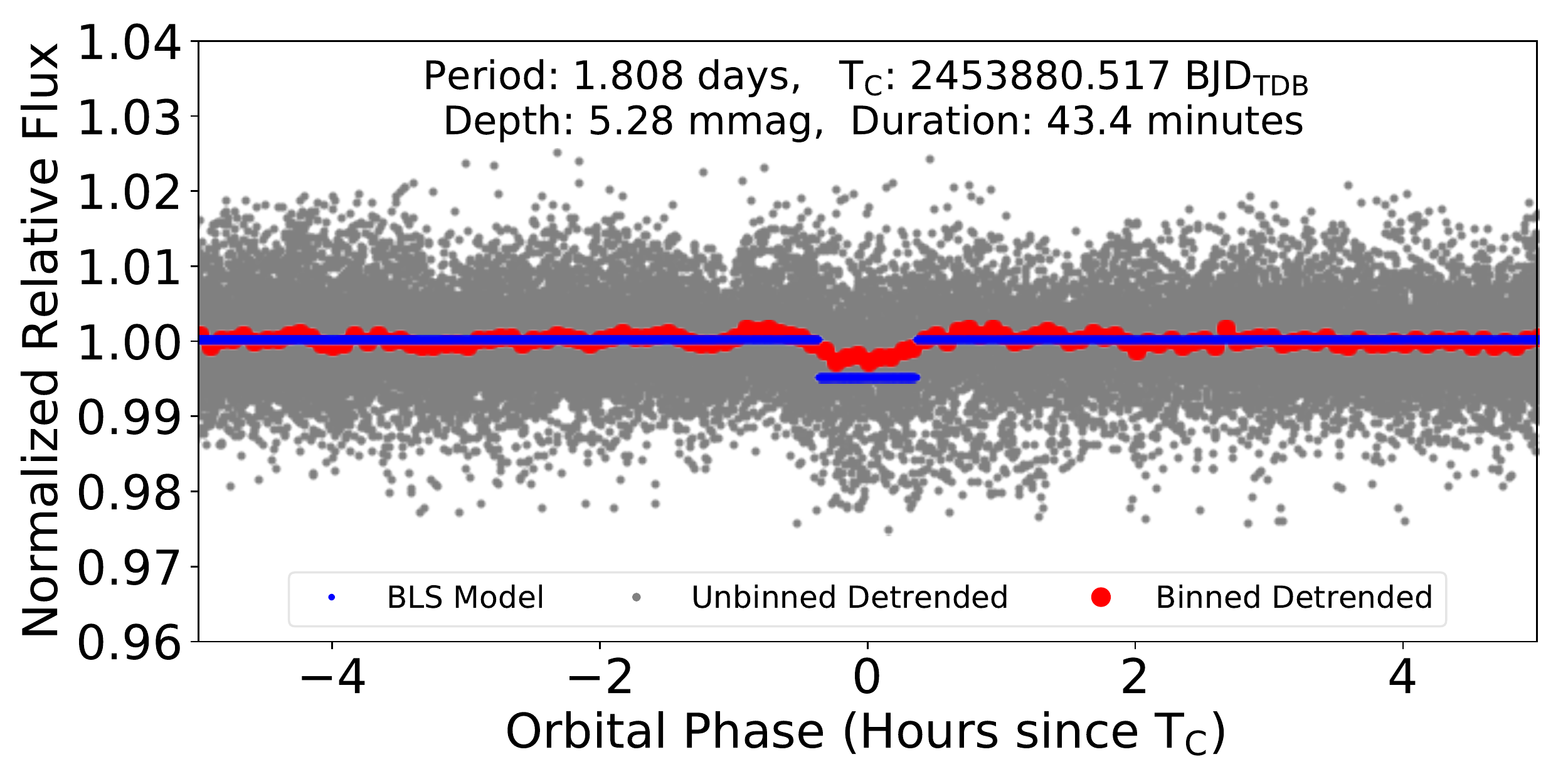}
\caption{From the power spectrum in Figure \ref{fig:BLS_1_30}, we took the orbital period corresponding to the top peak and phase folded our combined 262 light curves. The grey dots are the unbinned detrended data and the red dots are binned detrended data with 5 minute bin widths. The BLS model of this peak, shown as the blue line, has a transit depth and duration of $\sim5.28$ mmag and 43.4 minutes. We do not believe this to be a real transit event as shown in Figure \ref{fig:PFLCs_topBLSPeak} and explained in the Appendix section.}
\label{fig:BLS_1_30_top_peak}
\end{figure}

We note that there is no significant power in the BLS power spectrum at the orbital period determined by the radial velocity of Proxima b \citep{Anglada:2016} which is consistent with our failure to find transits in Paper I. In section \ref{sec:Huber}, we describe a methodology to detect low power peaks like those near the 11.186 day RV period that are displayed in the inset panel of Figure \ref{fig:BLS_1_30}.

\subsection{Testing Statistical Significance of BLS Power Spectra}\label{sec:FAP_tests}
\subsubsection{FAP}\label{sec:FAP_2}

In order for each power spectrum of the randomly permuted data sets to be on comparable scales, we normalized the power spectra using a modified version of the Spectroscopic Signal to Noise used in \citet{Hartman:2016}:

\begin{equation}\label{eq:SN}
    S/N(f) = \frac{ SR(f)_{random} - \overline{SR}(f) }{\sigma_{\overline{SR}}}
\end{equation}
\noindent
where we use the average and standard deviations of the BLS reported  \textit{SR(f)} of the power spectrum in Figure \ref{fig:BLS_1_30} with the \textit{SR(f)} of the randomized data sets. Based on our definition of FAP in Section \ref{sec:FAP}, we find FAP thresholds for the 0.1\% FAP occurs at a S/N of $\sim 7.61$, 1\% FAP at $\sim 6.96$ and the 10\% FAP at $\sim 6.35$ as shown as horizontal black lines in Figure \ref{fig:BLS_1_30}. 

\subsubsection{FAP As A Function of Period}\label{sec:FAP_P}
To better assess the validity of peaks below the 0.1\%, 1\% and 10\% FAP values, we divided $N_{freq}$ into 20 period ranges, starting from 1.01 days and used Equations \ref{eq:Nfreq} and \ref{eq:df} to calculate the bounds of each subsequent period range so that each range contained the same number of frequencies.

\begin{equation}\label{eq:PeriodRanges}
    \resizebox{0.42\textwidth}{!}{$P[i] = \frac{P[i-1]}{1-P[i-1] \times (q_{min}/4T)  \times (N_{freq}/20)}  \forall i \in  \{1,...,20 \}$}
\end{equation}

\noindent

\begin{figure}[htb] 

\centering %
\includegraphics[width=\columnwidth]{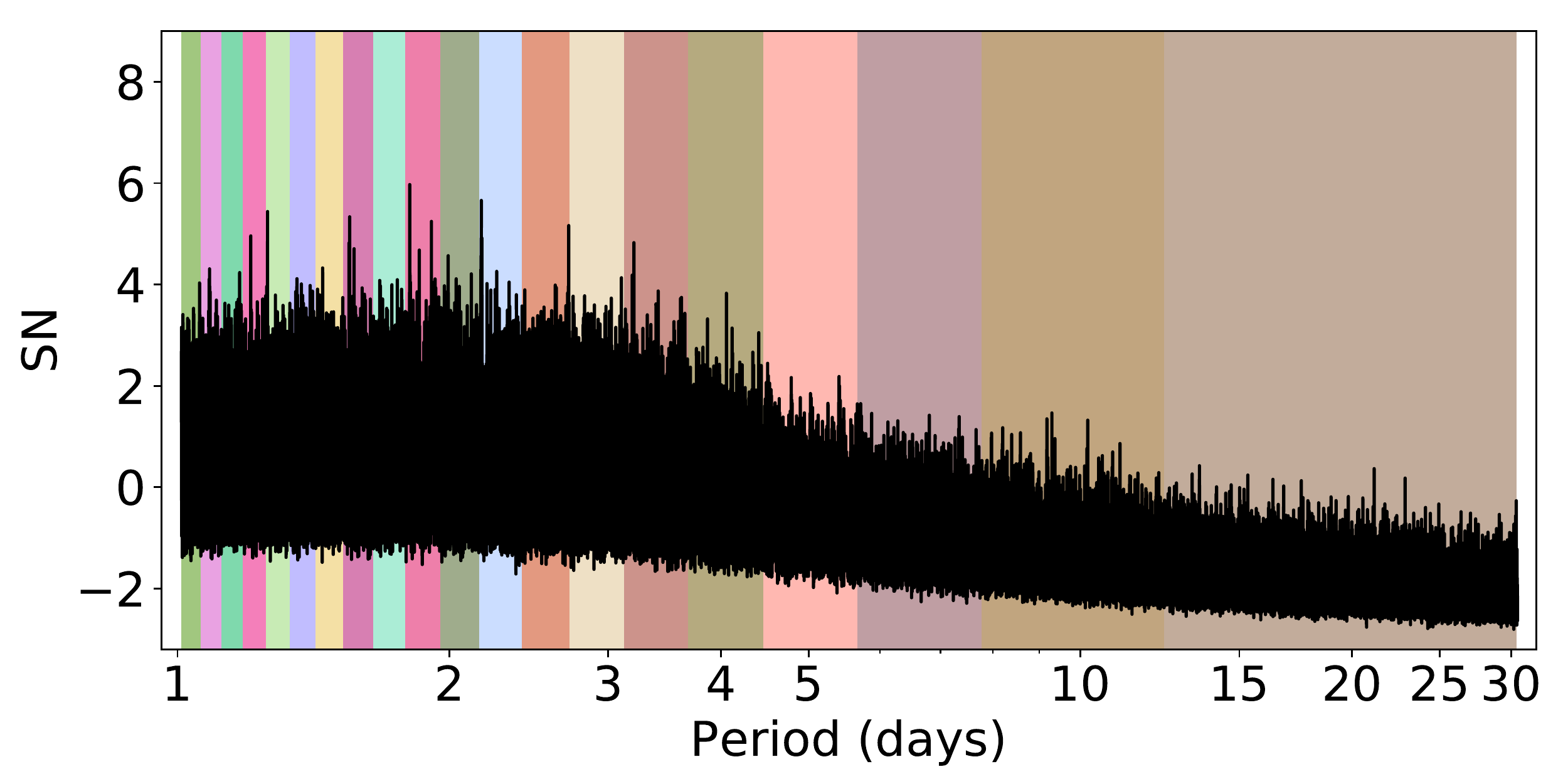}
\caption{An illustration of the 20 period ranges used to define FAP(P). The edges of the ranges are calculated with Equation \ref{eq:PeriodRanges}. Each colored shaded region is a different period range where each range has an even amount of frequencies, $N_{freq}/20$.}
\label{fig:BLS_PeriodRanges}
\end{figure}

\noindent

 These period ranges are displayed in Figure \ref{fig:BLS_PeriodRanges}. For each period range, we then follow a similar process as our FAP procedure, where we randomly shuffle our data with fixed time stamps and then run a BLS search on the shuffled data. Within each period range, we then record information from the highest peak in the resulting BLS power spectra and repeat the process a total of 1,000 times. From these 1,000 permutations, we calculate the FAP thresholds within each period range. We then interpolated FAP values between edges of each period range which we refer to hereafter as FAP(P), which is shown as the red line-connected dots in Figure \ref{fig:BLS_1_30}.

\subsubsection{Robust Estimation Of The Mean S/N}\label{sec:Huber}
\begin{figure*}
    \centering
    \includegraphics[scale=0.49]{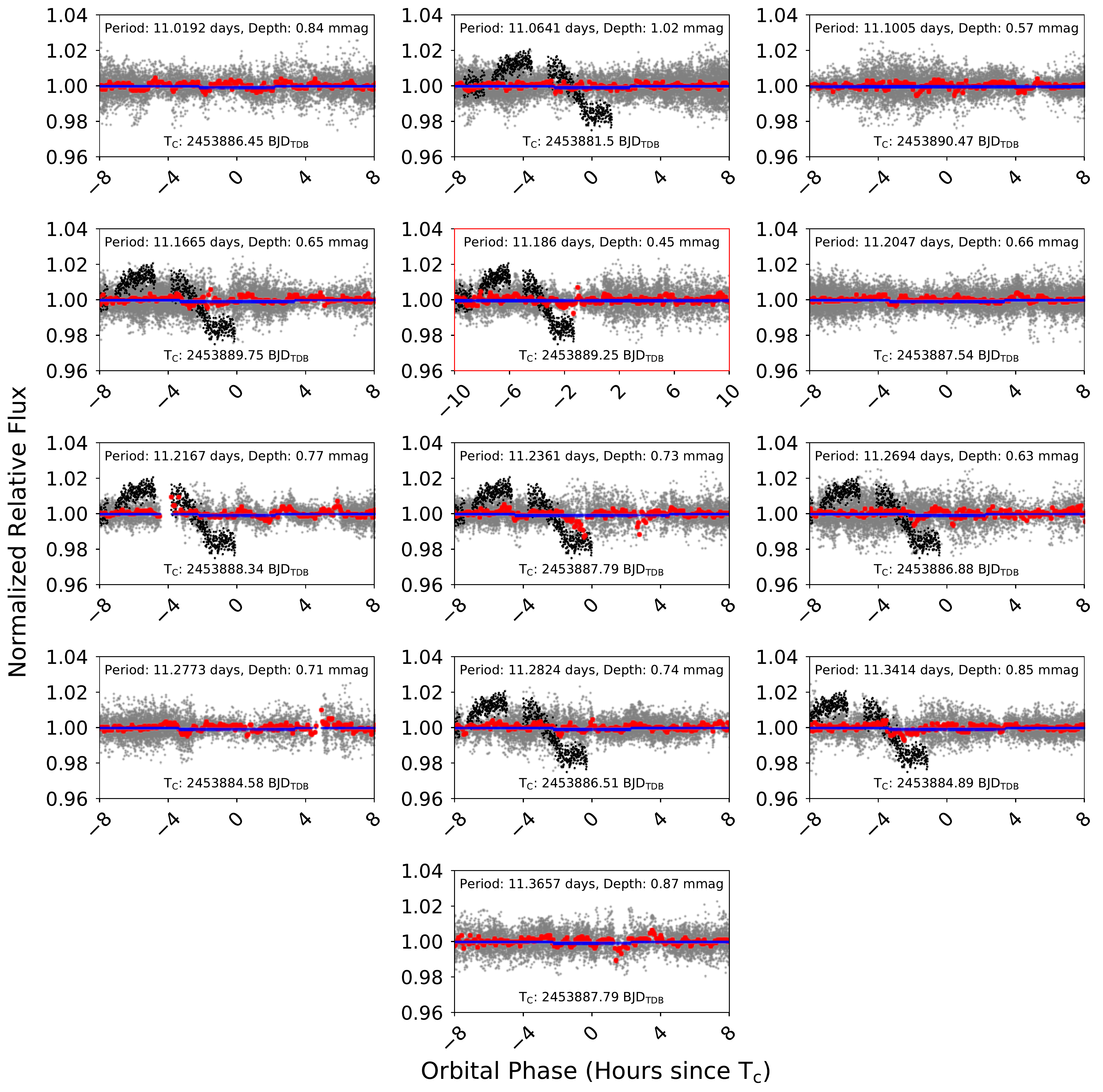}
\caption{From the inset panel of Figure \ref{fig:BLS_1_30}, we have identified 12 peaks that have S/N values larger than the robust mean + $3\sigma$ of the power spectrum's S/N within the corresponding period range. We then conducted a BLS search using a fixed period for each of these 12 peaks, in addition to the 11.186 day RV period, and phase folded our data with the corresponding transit center time output by the VARTOOLS BLS algorithm. In each panel of this figure, we have the phase folded light curves with the unbinned detrended data as grey points, the binned detrended data with 5 minute bins as red points and the BLS model as the blue lines. The black points are the detrended data from the UT April 11 2007 RAE light curve which is also in shown Figures \ref{fig:PFLCs_topBLSPeak} and discussed in the Appendix section. In each case, the transit depths reported by BLS are on the order of 1 mmag or smaller which is reflected by the relatively weak peak strength as seen in Figure \ref{fig:BLS_1_30}.}
\label{fig:PFLCs_nearRV}
\end{figure*}

As a visually intuitive alternative to identifying potentially significant, low power, peaks like those that are near the RV orbital period of $11.186$ days, we utilized the Python package StatsModels\footnote{https://www.statsmodels.org/stable/index.html\label{ftn:huber}} module for Huber's robust estimator of scale and location \citep{Huber} to estimate the mean and standard deviation of the S/N of the power spectrum. By fixing the orbital period on these 12 low power peaks near the 11.186 day period, we then ran a BLS search to obtain parameters for transit center time, transit duration and transit depth. In Figure \ref{fig:PFLCs_nearRV}, we use these parameters to phase fold our data around these 12 orbital periods corresponding to peaks near the 11.186 day RV period, in addition to the 11.186 day RV period itself. We then carefully and critically examined curves all light curves that contribute to each of these 13 peaks. We find that on average, these peaks correspond to transit depths of $\sim 0.73$ mmag and there are no consistent light curve events that display these periodic decreases in flux. 

To apply this method as an additional transit detection criteria in Section \ref{sec:TrasitInjectionRecovery}, we calculated the robust estimations of mean and standard deviation of the transit injected power spectra S/N within each of the 20 defined period ranges described in Section \ref{sec:FAP_P} and shown in Figure \ref{fig:BLS_PeriodRanges}. Similarly to our procedure in estimating our FAP(P) function in Section \ref{sec:FAP_P}, we then interpolated values of the robust statistics between edges of each period range to use in our transit injection recovery described in Section \ref{sec:TrasitInjectionRecovery}. Using these robust statistics provides a better estimation for the location of the mean and standard deviation of S/N without rejecting outliers. 

\vspace{1cm}
\subsection{Transit Injection Recovery}\label{sec:TrasitInjectionRecovery}
To test the sensitivity of the BLS algorithm's transit detection ability on our data set, we injected 550 transit models into our detrended data as described in Section \ref{sec:TransitInjections}. We then ran the BLS algorithm using identical input parameters as in Table \ref{tbl:BLSparam} and recorded the resulting power spectra and BLS reported transit parameters. In this analysis, a successfully recovered transit injection is defined as a peak in the power spectrum that is within $\pm~1\%$ of the injected transit model's orbital period and has a BLS Power above a detection threshold. We use the FAP, FAP(P) and robust estimations of the mean plus standard deviation of the S/N as three separate thresholds to gauge our ability to detect varying peak strengths in the transit injected power spectra. 

\begin{figure}[htb] 
\centering %
\includegraphics[width=\columnwidth]{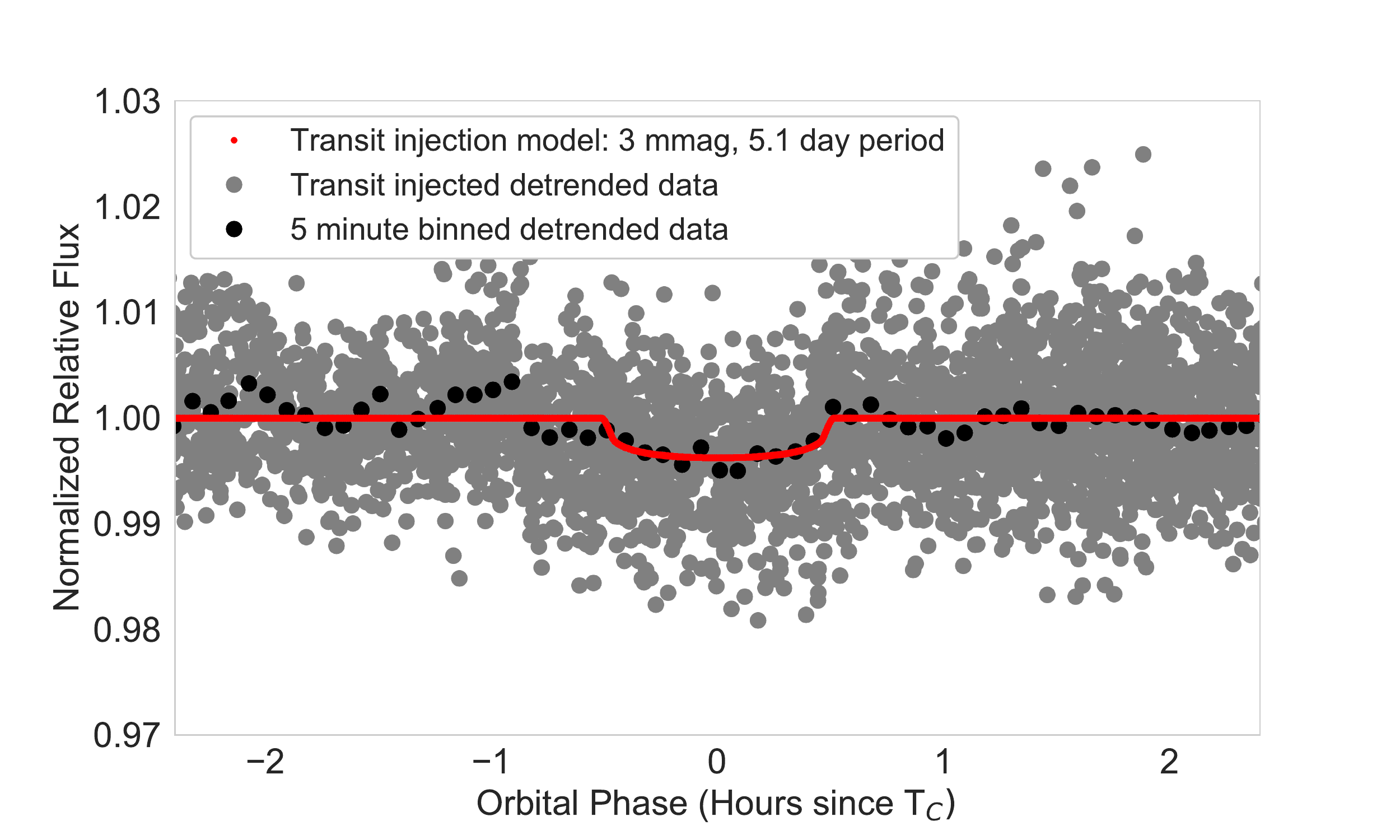}
\caption{With the Pytransit \citep{Parviainen:2015} python pacakge, we were able to inject 550 Mandel-Agol \citep{Mandel:2002} planet models into our detrended data. The model is displayed as a red line for one such simulated transiting planet with an orbital period of 5.1 days, a transit depth of 3 mmag and phase folded to the midpoint of our data set at 2455922.515188 \bjdtdb with an orbital phase of 0. The grey points are our transit injected detrended data. The black points are our transit injected detrended data binned with 5 minute bins.} 
\label{fig:transitinjection}
\end{figure}

We also considered harmonics and sub-harmonics (1/3, 1/2, 2 and 3 times) of the injected transit model periods in our detection criteria. We apply our detection criteria to the 550 transit injections and perform BLS searches to recover the injected transits across multiple orbital phases. As an example, Figure \ref{fig:transitinjection} displays 1 of our 550 transit injections that was successfully recovered. The results of using these three different thresholds along with our detection criteria can be seen in Figures \ref{fig:Global_FAP_Plots}, \ref{fig:FAPP_Plots} and \ref{fig:RobustStats_Plots}. The numbers of the color bars shown in Figures \ref{fig:Global_FAP_Plots}, \ref{fig:FAPP_Plots} and \ref{fig:RobustStats_Plots} represent the number of orbital phases where a detection by BLS occurred, ranging from 0 (for no detections in any of the phases tested) to 5 (detections in all phases tested) for the FAP, FAP(P) and robust mean plus standard deviation thresholds, respectively.

As expected, the number of successfully recovered transit injections decrease for orbital periods beyond 15 days where our phase coverage decreases to about 90\% as shown in Figure \ref{fig:phase_coverage}. For the injected transits with Proxima b's orbital period of 11.186 days and a transit depth of 5 mmag (highlighted by red and cyan colored boxes in Figures \ref{fig:Global_FAP_Plots}, \ref{fig:FAPP_Plots} and \ref{fig:RobustStats_Plots}) and higher, we successfully recover transit injections in at least 2 of the 5 orbital phases with and without the requirement of FAP or FAP(P), and recover 5 out of the 5 orbital phases with the robust mean plus standard deviation thresholds. 

\subsection{Constraints on Transiting Planet Properties}\label{sec:RV_model}
As a simple exercise, we also estimated which combinations of transit depth and orbital period could be detected by the Doppler semi-amplitude of Proxima b ($\sim 1.4$ m/s) as reported by \citet{Anglada:2016}. For a given planet radius ($R_p = \sqrt{\delta}R_*$), we can define the planet mass as:

\begin{equation}
    M_p = \bigg(\frac{\rho_p}{\rho_{\Earth}} \bigg){\bigg(
    \frac{R_p}{\RE}\bigg)}^{3} \ME
\end{equation}

\begin{figure*}[htp] 
    \centering
    \includegraphics[scale=0.47]{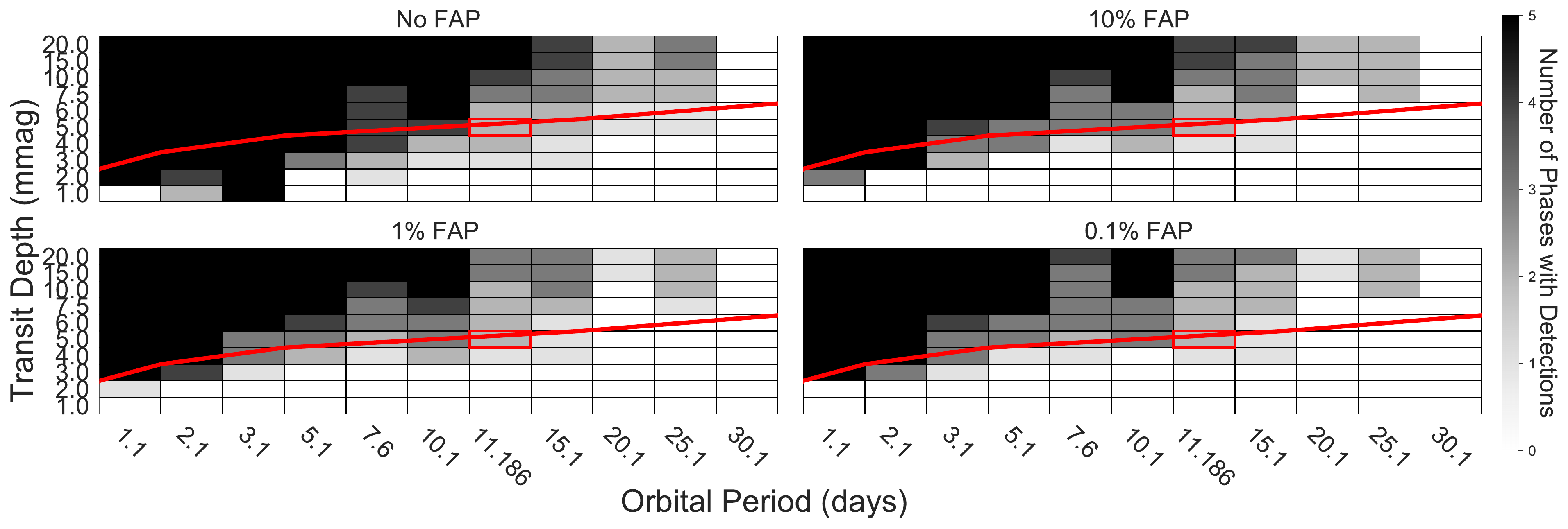}
    \caption{Applying the detection criteria described in Section \ref{sec:TrasitInjectionRecovery} to our 550 transit injected data sets, we present a color map of transit detections that occurred in the orbital phases: -0.4, -0.2, 0 , 0.2, 0.4. The upper left figure represents recovered transit injections within $\pm~1\%$ of their injected periods but considering no FAP threshold. The upper right figure represents recovered transit injections within $\pm~1\%$ of their injected periods and over the $10\%$ FAP thresholds. Similarly the lower left and right figures represent recovered transit injections within $\pm~1\%$ of their injected periods and above the $1\%$, $0.1\%$ FAP thresholds, respectively. The red line is our radial velocity model for an Earth-like exoplanet with constant Doppler semi-amplitude of 1.4 m/s, described in Section \ref{sec:RV_model}.}
    \label{fig:Global_FAP_Plots}
\end{figure*}

\begin{figure*}[htp] 
    \centering
    \includegraphics[scale=0.47]{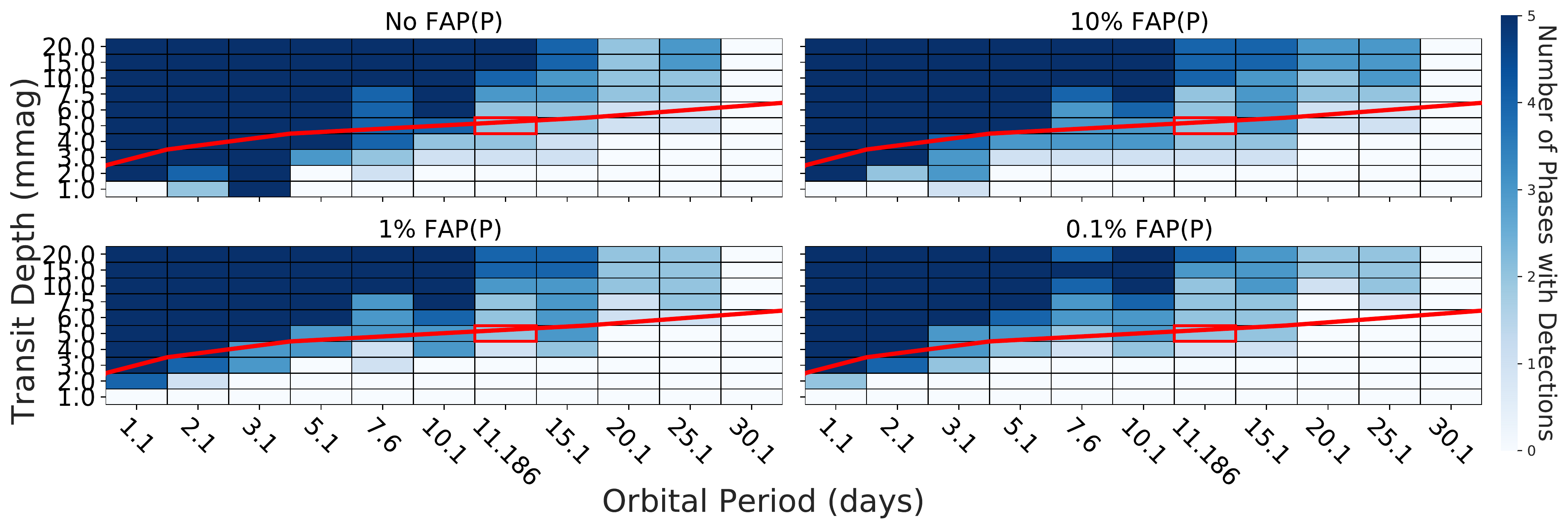}
    \caption{Similarly to Figure \ref{fig:Global_FAP_Plots}, we altered our detection criteria to use the FAP(P) function described in Section \ref{sec:FAP_P}. The upper left figure represents recovered transit injections within $\pm~1\%$ of their injected periods but considering no FAP(P) threshold. The upper right figure represents recovered transit injections within $\pm~1\%$ of their injected period and over the $10\%$ FAP(P) thresholds. Similarly the lower left and right figures represent recovered transit injections within $\pm~1\%$ of their injected period and above the $1\%$, $0.1\%$ FAP(P) thresholds, respectively. The red line is our radial velocity model for an Earth-like exoplanet with constant Doppler semi-amplitude of 1.4 m/s, described in Section \ref{sec:RV_model}.}
    \label{fig:FAPP_Plots}
\end{figure*}

\begin{figure*}[htb]
    \centering
    \includegraphics[scale=0.47]{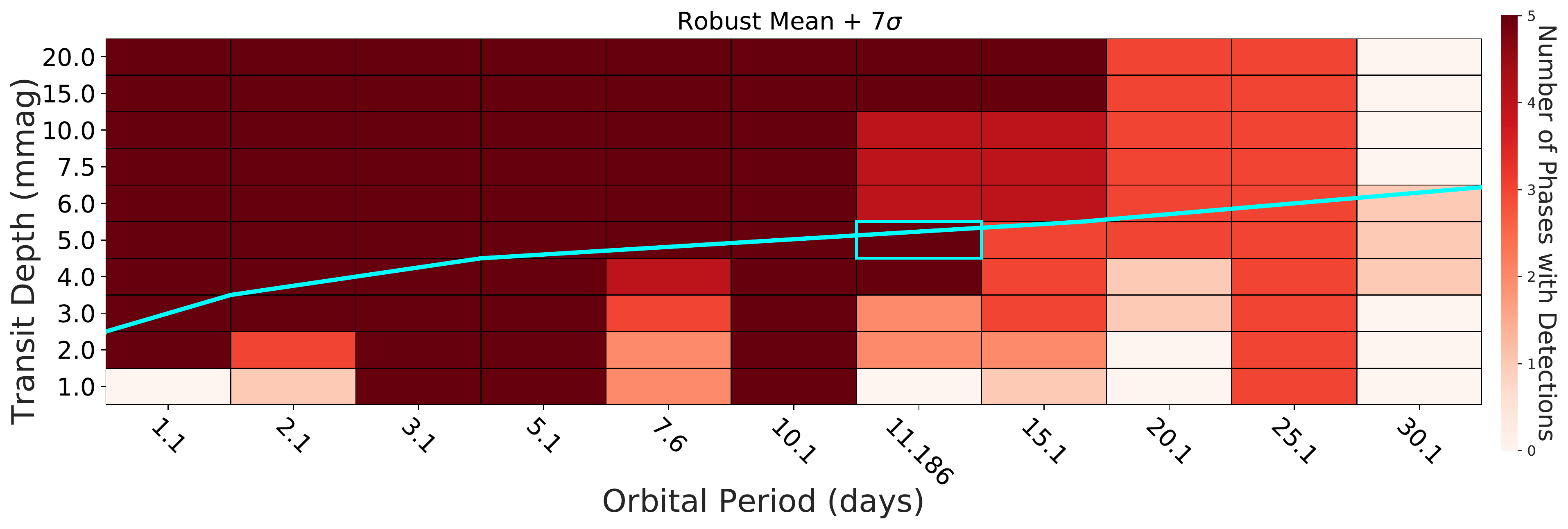}
    \caption{Similarly to Figure \ref{fig:Global_FAP_Plots}, we altered our detection criteria to use the robust estimation of mean, $\sigma$ of the BLS power spectrum as described in Section \ref{sec:Huber}. The recovered transit injections that are within $\pm~1\%$ of their injected periods and are over the robust mean + 7$\sigma$ of the injected power spectra. An example of the robust mean of a power spectrum can be seen as the green line in Figure \ref{fig:BLS_1_30}. The cyan line is our radial velocity model for an Earth-like exoplanet with constant Doppler semi-amplitude of 1.4 m/s, described in Section \ref{sec:RV_model}.}
    \label{fig:RobustStats_Plots}
\end{figure*}

\noindent
We assume the planet density to be Earth-like ($\rho_p / \rho_{\Earth} = 1$). By keeping the Doppler semi-amplitude (K) fixed, 
\begin{equation}
    K = \frac{2\pi~a\sin{i}}{P}
\end{equation}
applying Kepler's Third Law, and assuming a circular orbit (e=0), we can solve for the orbital period:

\begin{equation}\label{eq:RV_Period}
    P = 2\pi G {\bigg(\frac{M_p sin(i)}{K}\bigg)}^3 {\bigg(\frac{1}{M_*}\bigg)}^2
\end{equation}
\noindent
The RV model roughly follows our transit detection recoverability up to an orbital period $\sim 25$ days and transit depths as low as $\sim 6$ mmag. We display this model over our transit injection recovery plots as red and cyan colored lines in Figures \ref{fig:Global_FAP_Plots}, \ref{fig:FAPP_Plots} and \ref{fig:RobustStats_Plots}. Each orbital period and transit depth cell in Figures \ref{fig:Global_FAP_Plots}, \ref{fig:FAPP_Plots} and \ref{fig:RobustStats_Plots} that are above the line correspond to the transit depth and orbital period combinations that should be detectable in the RV data. 

If Proxima b does transit and is not denser than Earth, then the planet would appear in a cell above the RV model; however our light curve data rule out any transit above the curve for orbital periods shorter than $\sim 15$ days. For other rocky planets (with Earth-like density) that might exist in the system, they must be below the RV model in Figures \ref{fig:Global_FAP_Plots}, \ref{fig:FAPP_Plots} and \ref{fig:RobustStats_Plots} or else their RV signatures would presumably have been detectable. While our light curve data does not generally probe the regions under the RV model, we are able to rule out detectable transit events for orbital periods $\lesssim 5$ days and transit depths $\gtrsim 3$ mmag.

\section{Discussion}\label{sec:discussion}
Although we find no evidence for transits of Proxima b in the BLS analysis of our data, we cannot confidently rule out transits at the period of Proxima b due to incomplete orbital phase coverage and a lack of sensitivity to transits shallower than 4 mmag. However, we are able to virtually rule out any other unknown planet transits of Proxima with orbital periods shorter than 5 days and depths greater than 3 mmag. Furthermore, within our phase coverage and depth sensitivity limitations (see Section \ref{sec:results}), we find no evidence for transits in our data for orbital periods in the range of 1 to 30 days.

In Paper I, we describe the selection criteria based on the amount of scatter in individual light curves that led to 262 of our 329 light curves to be included in our overall analysis. As shown in Figure 1 of \citet{Blank:2018}, the median standard deviation of individual unbinned light curves after the detrended and vetting processes is $\sim 0.52\%$ which is similar to the expected 0.5\% transit depth for Proxima b. For the minimum mass of 1.27 \ME~estimated by \citet{Anglada:2016}, smaller radii of Proxima b would translate to higher planet densities. 

\citet{Brugger:2016} determine the ranges of planet mass and radius of Proxima b to be (1.1 -1.46) \ME~and (0.94 - 1.40) \RE. In the case of the 1.1 \ME, 0.94 \RE~model of Proxima b, 65\% of the planet’s mass is located in the core and the remaining 35\% as part of the mantle. In the 1.46 \ME, 1.4 \RE~case, the corresponding composition is 50\% of the planet’s mass being in the form of water and the remaining 50\% in the mantle. We estimate a planet radius and mass of $\sim 0.94$ \RE~and $\sim 1.1$ \ME~corresponds to a transit depth of about 3 mmag and a planet density $1.32~ \rho_{\oplus}$ while a planet of radius 1.4 \RE~and mass 1.46 \ME~corresponds to a transit depth of about 7 mmag and a planet density of $\sim 0.5~ \rho_{\oplus}$.

To determine our lower limit for detectable planet densities, we have extended our exercise Section \ref{sec:RV_model} to also fix the planet mass to 1.27 \ME~in addition to the fixed Doppler semi-amplitude of 1.4 m/s. Using Equation \ref{eq:RV_Period} gives an orbital period $\sim$13.09 days. The lowest recovered injected transit depth near that orbital period in Figures \ref{fig:Global_FAP_Plots} and \ref{fig:FAPP_Plots} is $\sim$ 3 mmag which corresponds to a planet radius $\sim$0.92 \RE. This results in a planet density of $\sim$1.63 $\rho_{\oplus}$ which is below the minimum density of $\sim$2.07 $\rho_{\oplus}$ estimated from the lower bounds of the probabilistically constrained result from \citet{Bixel:2017} for their rocky planet model.

\citet{Loyd:2018} discusses the mechanisms of flares potentially inducing photoionization heating of the upper atmosphere of planets orbiting very near their host stars. Through the authors' work, they determined that the extreme ultraviolet radiation during flare events can be intense enough to drive hydrodynamic escape. \citet{Howard:2018} detected a super flare where Proxima's optical flux increased by 68 times in an hour long event. Through the two years of observations with the Evryscope telescope, they observed 23 other large flare events and determined that super flares of this scale occur roughly 5 times per year, and that this level of repeated flaring may be sufficient enough to reduce the ozone of an Earth-like atmosphere by 90\% within five years and complete depletion may occur within several hundred thousand years. In the event of Proxima b's planetary radius being smaller due to a different planet density or decreasing due to flare driven atmospheric loss, our data should have a sensitivity to transit depths up to $\sim 0.5\%$, supported by the 5 minute binned RMS of our data being $\sim~0.26\%$.

There are several areas for improvement within our data reduction pipeline. The main challenge of detrending our ground-based observations was obtaining as flat of a photometric baseline as possible. With the majority of our light curves coming from unguided telescopes, there existed discontinuous jumps in the raw data from meridian flips and telescope re-pointings. Our iterative 3-$\sigma$ clipping of Proxima's frequent flare events followed by our detrending methodology did not completely flatten the baseline of light curves and may provide some periodic power in the BLS power spectrum. Figure \ref{fig:PFLCs_topBLSPeak} has a few examples of this such as the UT June 17, 2014 Prompt 1 and Prompt 4 light curves. Algorithms like BLS could fit a box model to these discontinuities or post $3\sigma$-cut flare remnants and report an unlikely transit event.

In our transit injection analysis in Section \ref{sec:TransitInjections}, we injected our simulated planet models into our detrended data. With our current machinery, it is impractical to inject our 550 planet models and then detrend each injected data set individually with AstroImageJ. In our upcoming Paper III (D.L. Feliz et al., in preparation) we will approach our transit injection methodology to incorporate detrending after injection of simulated planet models.

It is likely that our data have residual correlated noise after our detrending process and there are a many ways to approach modeling the noise (e.g. Gaussian Processes). In addition to scatter due to the intrinsic variability of the star, Proxima is also a well known flare star \citep{Shapley:1951, Walker:1981}. There are numerous flares within our data set that can be quantified (Jayawardene et al., in preparation) and modelled to subtract the events from our light curves to obtain a flatter baseline more suitable for transit detection. Another alternative to estimate the FAP is with Bayesian statistics rather than our Frequentist approach which assumes pure white noise.
\vspace{1mm}

\section{Conclusion\label{sec:conclusion}}
In this work, we present an analysis of 262 photometric observations of Proxima Centauri where we search for periodic transit-like events. The light curves have been cleaned and detrended as described in Paper I, and then fed into the box-fitting least squares (BLS) period finding algorithm. To estimate the statistical significance of the peaks in the BLS power spectrum, we estimated 0.1\%, 1\% and 10\% false alarm probability (FAP) thresholds. We also determined FAP as a function of orbital period (FAP(P)) by calculating the FAP thresholds for 20 period ranges, each with an equal number of frequencies to be searched for by the BLS algorithm. To explore peaks in the BLS power spectra that have low power, we used Huber's robust estimator of scale and location to estimate the mean and standard deviation of the power spectrum in each of the 20 period ranges. The majority of the highest peaks of the BLS power spectrum fall below the FAP and FAP(P) thresholds. We note that there is no significant power in the power spectrum near the orbital period of Proxima b \citep{Anglada:2016}; however we have identified 12 peaks that are above the robust mean plus 3 and 5 times the standard deviation. We then phase folded those 12 peaks and examined individual light curves that contribute to those periods. We conclude that these 12 peaks are unlikely to be caused by transit events. 

To test our sensitivity for detecting transit-like events, we injected 550 fake transits with parameters differing in transit depth, orbital phase and orbital period. We were able to detect most injections at transit depths 4 mmag and greater up to an orbital period of 11.186 days across multiple orbital phases. Overall, we were unable to confirm the existence of transits of Proxima b. 
\clearpage
In our upcoming Paper III (D.L. Feliz et al., in preparation), we intend to model the numerous flares in our data as well as the correlated noise to reduce the scatter in our light curves and conduct a more thorough period finding search. \\

\noindent
\textit{Software Used:\\} AstroImageJ \citep{Collins:2017}, EXOFAST website\footnotemark[3] \citep{Eastman:2013}, VARTOOLS \citep{Hartman:2016}, and PyTransit \citep{Parviainen:2015}. 

\vspace{3 cm}
\acknowledgements
The authors thank the anonymous referee for helpful suggestions that improved this manuscript. The authors would also like to thank Joshua Pepper, Scott Gaudi, David Latham, Michael Lund, and Robert Siverd for their thoughtful discussions and input regarding our data reduction and analysis. Additionally, the authors would like to once again thank the science teams of RAE, Skynet and KELT-FUN for their contributions to our survey. U.S. Naval Observatory astronomy is supported by Navy R\&D and Operational laboratory funding. Dax L. Feliz would also like to acknowledge support from the Fisk-Vanderbilt Masters-to-PhD Bridge Program.\\
\bibliographystyle{apj}

\bibliography{main}

\begin{figure}
    \centering
    \includegraphics[scale=0.25]{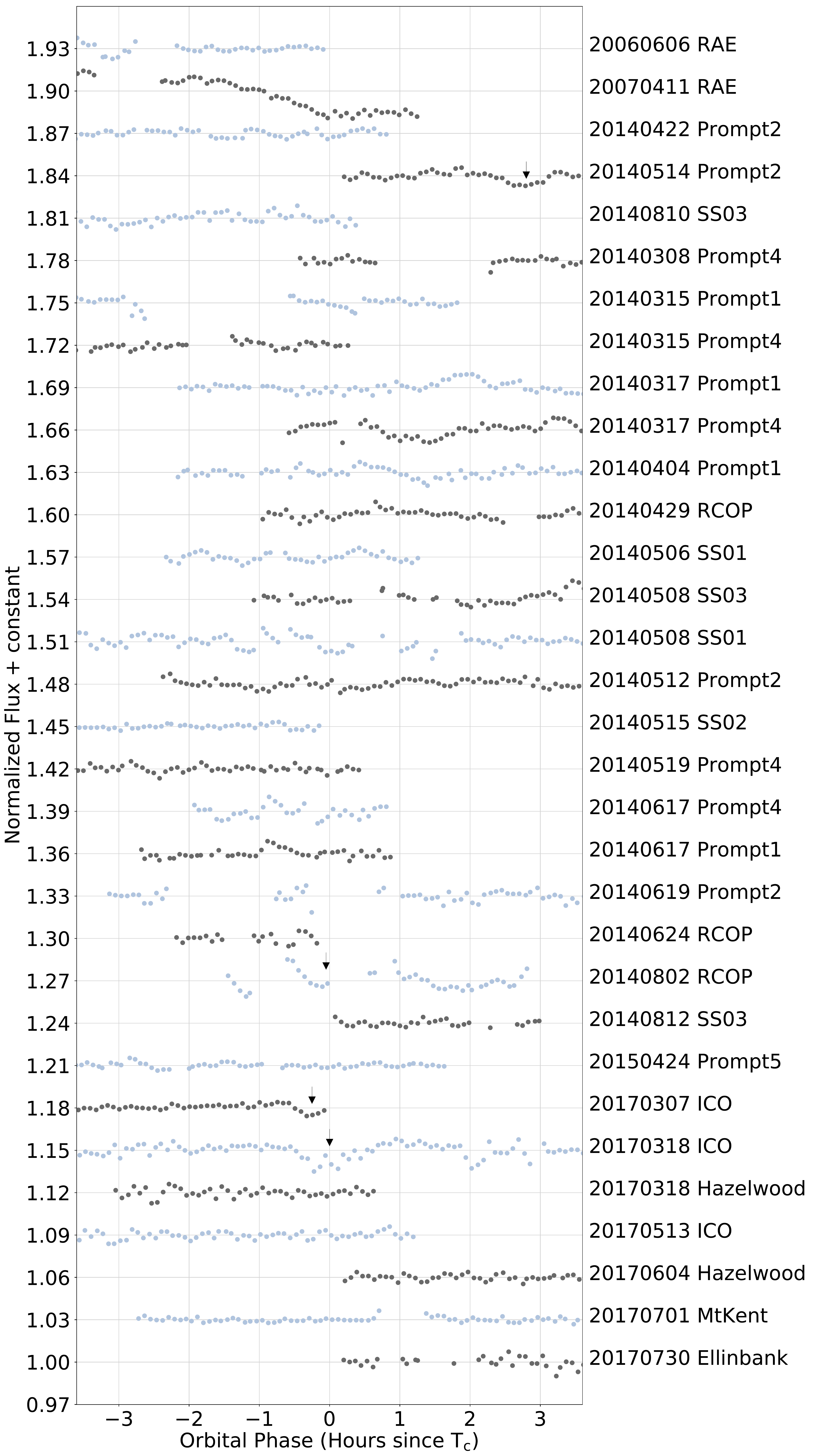}
\caption{These detrended light curves correspond to the contributions in phase of Figure \ref{fig:BLS_1_30_top_peak}. Each light curve is phase folded around the orbital period and transit center time from the highest peak from the VARTOOLS BLS algorithm. The orbital period for this peak is 1.808 days and has a transit depth of 5.28 mmag. We vertically separated each light curves by a constant and alternated their colors for easier distinction between observation. Each light curve is binned at 5 minute intervals and contributes at least one data point within $\pm 1$ hour of the transit center time. We discuss the light curves highlighted by arrows in more detail in the Appendix section.}
\label{fig:PFLCs_topBLSPeak}
\end{figure}

\appendix \label{sec:appendix}
To verify whether or not the top peak of the BLS power spectrum is due to a transit-like event, we phase folded our 262 light curves around the orbital period corresponding to the top peak which is $\sim1.808$ days. We then examined a subset of 32 light curves that contribute data points to within $\pm 3$ hours of the transit center time reported by the BLS algorithm, $T_C \sim$ = 2453880.516604 \bjdtdb. In Figure \ref{fig:PFLCs_topBLSPeak}, the individual light curves are then binned with 5 minute bins and are vertically offset from one another. In Figure 11 of Paper I, we modeled the best fit transit model of the Skynet Prompt 2 UT 2014 May 14 light curve where we deemed it unlikely to be caused by a transiting exoplanet. The Skynet RCOP light curve from UT 2014 August 2 has a decrease in flux near phase 0 relative to $T_C$ but this is due to the remaining points after our iterative 3-$\sigma$ cut of a flare event that is followed by another smaller pair of flare events. The KELT-FUN Ivan Curtis Observatory light curves from UT 2017 March 7 and UT 2017 March 18 display decreases in flux near phase 0 in relation to $T_C$. In the UT 2017 March 7 light curve, the decrease at the end of the light curve are suspected to be due to atmospheric fluctuation but remained in our quality checks in Paper I. The UT March 18 2017 has a relatively high amount of scatter compared to the rest of our data but also passed our quality check due to the transit-like feature shown. In Figure \ref{fig:PFLCs_nearRV}, we highlighted data from the UT April 11 2007 RAE light curve which displays a $\sim$20 mmag dip which we've chosen to include in our ensemble of light curves. We conclude that the that top peak of the BLS power spectrum shown in Figure \ref{fig:BLS_1_30_top_peak} is unlikely to be due to a transiting exoplanet as there is no consistent transit events in all other light curves near orbital phase 0 relative to $T_C$.

\begin{table*}[htb!]
\renewcommand\thetable{A.1}
\begin{center}
\begin{threeparttable}
\caption{Photometry of Proxima Centauri\label{tbl:data1}}{\setlength{\tabcolsep}{1em}
\begin{tabular}{lcc}
\tableline

\multicolumn{1}{|c|}{\bjdtdb}& \multicolumn{1}{c}{Normalized Relative Flux}&\multicolumn{1}{|c|}{Uncertainty}    \\ \hline  

\multicolumn{1}{|c|}{2453879.916949} & \multicolumn{1}{c|}{1.03734}&\multicolumn{1}{|c|}{0.00461}    \\ 

\multicolumn{1}{|c|}{2453879.917331} & \multicolumn{1}{c|}{1.02921}&\multicolumn{1}{|c|}{0.00439}    \\ 

\multicolumn{1}{|c|}{2453879.917713} & \multicolumn{1}{c|}{1.02177}&\multicolumn{1}{|c|}{0.00427}    \\ 

\multicolumn{1}{|c|}{2453879.918095} & \multicolumn{1}{c|}{1.01875}&\multicolumn{1}{|c|}{0.00410}    \\ 

\multicolumn{1}{|c|}{2453879.918477} & \multicolumn{1}{c|}{1.01907}&\multicolumn{1}{|c|}{0.00405}    \\ 

\multicolumn{1}{|c|}{2453879.918870} & \multicolumn{1}{c|}{1.01457}&\multicolumn{1}{|c|}{0.00392}    \\ 

\multicolumn{1}{|c|}{\vdots} & \multicolumn{1}{c|}{\vdots}&\multicolumn{1}{|c|}{\vdots}    \\ 

\multicolumn{1}{|c|}{2457965.105301} & \multicolumn{1}{c|}{1.028180}&\multicolumn{1}{|c|}{0.004330}    \\ 

\multicolumn{1}{|c|}{2457965.108993} & \multicolumn{1}{c|}{1.032670}&\multicolumn{1}{|c|}{0.003000}    \\ 

\multicolumn{1}{|c|}{2457965.109398} & \multicolumn{1}{c|}{1.002570}&\multicolumn{1}{|c|}{0.004370}    \\ 

\multicolumn{1}{|c|}{2457965.109838} & \multicolumn{1}{c|}{1.038580}&\multicolumn{1}{|c|}{0.004730}    \\ 

\multicolumn{1}{|c|}{2457965.110706} & \multicolumn{1}{c|}{1.018970}&\multicolumn{1}{|c|}{0.003880}    \\ 

\multicolumn{1}{|c|}{2457965.111134} & \multicolumn{1}{c|}{1.035270}&\multicolumn{1}{|c|}{0.003390}    \\ \hline
\end{tabular}}

   \begin{tablenotes}[flushleft]
      \small
      \item This table contains photometry of the 262 combined light curves used in \citealt{Blank:2018} and this work. This data set is undetrended and iteratively 3-sigma clipped. The time stamps are the barycentric julian date in the barycentric dynamical time of observation (\bjdtdb). The data in its entirety is in the electronic version of \textit{The Astrophysical Journal}. A portion is shown here for guidance regarding its form and content.
    \end{tablenotes}
\end{threeparttable}
\end{center}
\end{table*}

\begin{table*}[htb!]
\renewcommand\thetable{A.2}
\begin{center}
\begin{threeparttable}
\caption{Detrended Photometry of Proxima Centauri\label{tbl:data2}}{\setlength{\tabcolsep}{1em}
\begin{tabular}{lcc}
\tableline

\multicolumn{1}{|c|}{\bjdtdb}& \multicolumn{1}{c}{Normalized Relative Flux}&\multicolumn{1}{|c|}{Uncertainty}    \\ \hline  

\multicolumn{1}{|c|}{2453879.928989} & \multicolumn{1}{c|}{1.005052}&\multicolumn{1}{|c|}{0.003787}    \\ 

\multicolumn{1}{|c|}{2453879.929371} & \multicolumn{1}{c|}{1.010987}&\multicolumn{1}{|c|}{0.003808}    \\ 

\multicolumn{1}{|c|}{2453879.929764} & \multicolumn{1}{c|}{1.010987}&\multicolumn{1}{|c|}{0.003740}    \\ 

\multicolumn{1}{|c|}{2453879.930135} & \multicolumn{1}{c|}{1.007096}&\multicolumn{1}{|c|}{0.003693}    \\ 

\multicolumn{1}{|c|}{2453879.930517} & \multicolumn{1}{c|}{1.002253}&\multicolumn{1}{|c|}{0.003618}    \\ 

\multicolumn{1}{|c|}{2453879.930899} & \multicolumn{1}{c|}{0.990985}&\multicolumn{1}{|c|}{0.003557}    \\ 

\multicolumn{1}{|c|}{\vdots} & \multicolumn{1}{c|}{\vdots}&\multicolumn{1}{|c|}{\vdots}    \\ 

\multicolumn{1}{|c|}{2457965.104885} & \multicolumn{1}{c|}{1.008081}&\multicolumn{1}{|c|}{0.004403}    \\ 

\multicolumn{1}{|c|}{2457965.105301} & \multicolumn{1}{c|}{1.003847}&\multicolumn{1}{|c|}{0.004225}    \\ 

\multicolumn{1}{|c|}{2457965.108993} & \multicolumn{1}{c|}{1.007049}&\multicolumn{1}{|c|}{0.002922}    \\ 

\multicolumn{1}{|c|}{2457965.109838} & \multicolumn{1}{c|}{1.012655}&\multicolumn{1}{|c|}{0.004612}    \\ 

\multicolumn{1}{|c|}{2457965.110706} & \multicolumn{1}{c|}{0.992760}&\multicolumn{1}{|c|}{0.003777}    \\ 

\multicolumn{1}{|c|}{2457965.111134} & \multicolumn{1}{c|}{1.008889}&\multicolumn{1}{|c|}{0.003305}    \\ \hline
\end{tabular}}

   \begin{tablenotes}[flushleft]
      \small
      \item This table contains photometry of the detrended 262 combined light curves used in \citealt{Blank:2018} and this work. This data set is processed as described in \citealt{Blank:2018}. The time stamps are the barycentric julian date in the barycentric dynamical time of observation (\bjdtdb). The data in its entirety is in the electronic version of \textit{The Astrophysical Journal}. A portion is shown here for guidance regarding its form and content.
    \end{tablenotes}
\end{threeparttable}
\end{center}
\end{table*}

\end{document}